\documentstyle[preprint,aps,psfig]{revtex}
\begin{document}
\draft
\title{Plasma heating due to X-B mode conversion in a cylindrical ECR
plasma system}
\author{Vipin K. Yadav\cite{mail1} and D. Bora\cite{mail2}}
\address{Institute for Plasma Research, Bhat, Gandhinagar, Gujarat -
382 428, India}
\maketitle
\begin{abstract}

\noindent
Extra Ordinary (X) mode conversion to Bernstein wave near Upper Hybrid
Resonance (UHR) layer plays an important role in plasma heating through
cyclotron resonance. Wave generation at UHR and parametric decay at
high power has been observed during Electron Cyclotron Resonance (ECR)
heating experiments in toroidal magnetic fusion devices. A small linear
system with ECR and UHR layer within the system has been used to conduct
experiments on X-B conversion and parametric decay process as a function
of system parameters. Direct probing {\em in situ} is conducted and
plasma heating is evidenced by soft x-ray emission measurement.
Experiments are performed with hydrogen plasma produced with 160-800 W
microwave power at 2.45 GHz of operating frequency at $10^{-3}$ mbar
pressure. The axial magnetic field required for ECR is such that the
resonant surface (B = 875 G) is situated at the geometrical axis of the
plasma system. Experimental results will be presented in the paper.

\end{abstract}
\section{Introduction}

\noindent
Electron cyclotron resonance heating (ECRH) is a highly efficient,
controllable and localized plasma heating \cite{lloyd} and current
drive scheme in fusion machines such as tokamaks
\cite{manheimer,england,eldridge,erckmann},
stellerators \cite{abrakov} and plasma systems such as tandem mirrors
\cite{porkolab}. These devices, when opearted at high plasma densities
(overdense plasma, $\omega_{pe}$ > $\omega_{ce}$), direct ECR heating
and current drive by the incident electromagnetic (EM) wave is not
possible. Alternatively, linear conversion of the EM waves into the
electrostatic (ES) EBW \cite{bernstein} at the UHR surface and
subsequent damping at the cyclotron harmonics is being tried for
overdense plasmas \cite{piliya}.

\noindent
In ECRH, the launched EM wave in extraordinary (X) mode, ($\vec k \perp
\vec B_0$ and $\vec E_1 \perp \vec B_0$, $\vec k$ is the wave vector,
$\vec B_0$ is the magnetic field in the system and $\vec E_1$ is the
wave electric field), is absorbed at the UHR surface in plasma as shown
in the X-mde dispersion relation
\begin{equation}
\frac {c^2k^2}{\omega^2} = 1 - \frac {\omega_{pe}^2}{\omega^2}\left(
\frac {\omega^2 - \omega_{pe}^2}{\omega^2 - \omega_{uh}^2}\right)
\label{ebw:eq1}
\end{equation}
where, $\omega$ is the launched microwave frequency, $\omega_{pe}$ is
the electron plasma frequency and $\omega_{uh}$ is the upper hybrid
frequency given by 
\begin{equation}
\omega_{uh}^2 = \omega_{pe}^2 + \omega_{ce}^2
\label{ebw:eq2}
\end{equation}
here, $\omega_{ce}$ is the electron cyclotron frequency.

\noindent
Near the UHR layer, the X-mode wavelength becomes very short and the
amplitude of wave electric field becomes very large (high
power-density). This large wave electric field gives rise to nonlinear
phenomena of parametric excitation in which the incoming EM wave breaks
up into EBW (X-B mode conversion) and lower hybrid (LH) wave
\cite{istomin}. The EBW propagates perpendicularly to the magnetic field
towards the plasma centre and gets strongly damped at ECR layer as shown
by the EBW dispersion relation given by
\begin{equation}
1 + \left(\frac {k_B v_{th}}{\omega _{pe}}\right)^2 =
e^{\left(-{k_B^2 r_L^2}\right)}I_0 \left({k_B^2}{r_L^2}\right) -
2\left(\frac{\omega}{\omega _{ce}^2}\right){\sum _q}
e^{\left(-{k_B^2}{r_L^2}\right)}{\frac
{I_q\left({k_B^2}{r_L^2}\right)}{\left(q^2 -
{\frac {\omega ^2}{\omega _{ce}^2}}\right)}}
\label{ebw:eq3}
\end{equation}
here, $v_{th}$ is the electron thermal velocity, $r_L$ is the electron
Larmour Radius and $I_q$ is the Bessel function of the first kind of
imaginary argument, q = 1, 2... The absorption of EBW contribute in
efficient plasma heating \cite{sugai,nakajima,ram}

\noindent
The parametric decay spectrum consists of wave with frequency of the
incident pump wave, $\omega$ and $\omega \pm \omega_{LH}$. The
characteristic sidebands in the high-frequency spectrum and the
associated low-frequency waves from the parametric excitation have been
observed in tokamaks \cite{mcdermott} and stellarators
\cite{laqua1,laqua2}.

\noindent
The EBW is observed experimentally in linear or toroidal plasma system
\cite{gekelman,nakajima}. The EBW generation via mode conversion is
observed in a large diameter linear plasma system with filament aided
discharge \cite{sugai}.

\section{Experimental set-up}

\begin{figure}[h]
\centerline{\hbox{\psfig{file=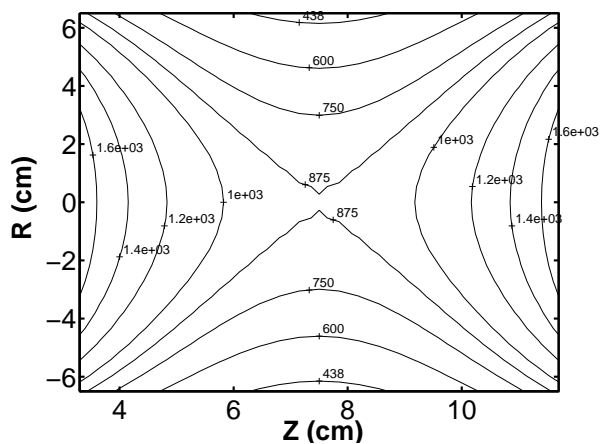,height=2.3in,angle=0}}}
\caption{Magnetic field contours in the system.}
\label{cont}
\end{figure}

\noindent
The experimental setup consists of a cylindrical stainless steel vacuum
chamber with internal radius 6.4 cm and axial length 10 cm. A base
pressure of $6 \times 10^{-6}$ mbar is achieved prior to the experiment
using a diffusion pump. The magnetic field is produced by two identical
circular coils rested on the axial ports at the two ends of the vacuum
vessel. Each magnetic field coil has an electrical resistance of 30
$\Omega$ and require 5.3 A of dc current to produce magnetic field of
875 G at the axis of the vacuum vessel which is kept constant for more
than 1 second. The magnetic field contours in the system are shown in
figure \ref{cont} \cite{vipin1}.

\noindent
The microwave system consists of a 800 W, $2.45 \pm 0.02$ GHz magnetron,
a three port circulator with water cooled dummy load to dissipate the
reflected power, WR340 directional coupler, low barrier Schottky diode
microwave detectors to measure forward and reflected power and a 10 mm
thick, $110 ~mm \times 70 ~mm$ rectangular toughen glass sheet to serve
as the waveguide window. In absence of plasma $\sim$ 80 \% of the
forward power gets reflected whereas the reflected power drops to
$\sim$ 20 \% of the forward power in presence of plasma.

\begin{figure}
\centerline{\hbox{\psfig{file=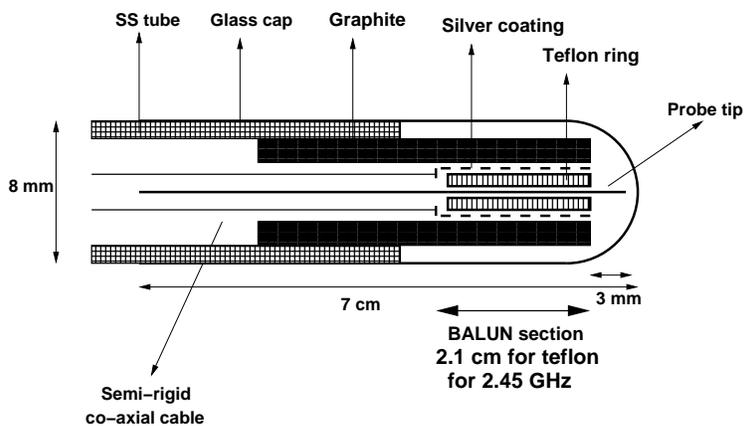,height=2.3in,angle=0}}}
\caption{Capacitive probe.}
\label{cprobe}
\end{figure}

\noindent
The plasma parameters are measured by a movable Langmuir probe
\cite{langmuir} in the radial direction. The probe is made of tungsten
with tip length and diameter of 5 mm and 0.5 mm respectively. A
capacitive probe \cite{vipin2} as shown in figure \ref{cprobe}, is used
to measure the wave field. Probe signal is directly analysed with a
spectrum analyser. Two capacitive probes \cite{vipin3} are used to
measure the wave characteristics having tip length less than the
expected wavelength to be detected, estimated theoretically as
$\approx$ 7.5 mm. The signals from the capacitive probes are fed into
two double balanced mixers (DBM). Each mixer has three ports : two
input - RF and LO (local oscillator) ports and one output - IF
(intermediate frequency).

\section{Experimental results}

\begin{figure}[h]
\centerline{\hbox{\psfig{file=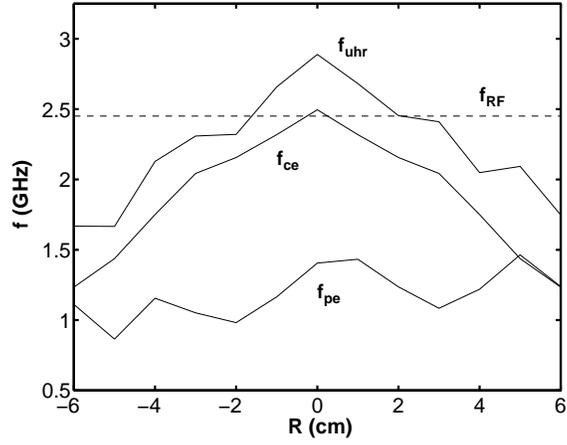,height=2.3in,angle=0}}}
\caption{Radial variation of $f_{ce}$, $f_{ce}$ and $f_{uhr}$ in
hydrogen plasma.}
\label{wcpuhrhy}
\end{figure}

\begin{figure}[h]
\centerline{\hbox{\psfig{file=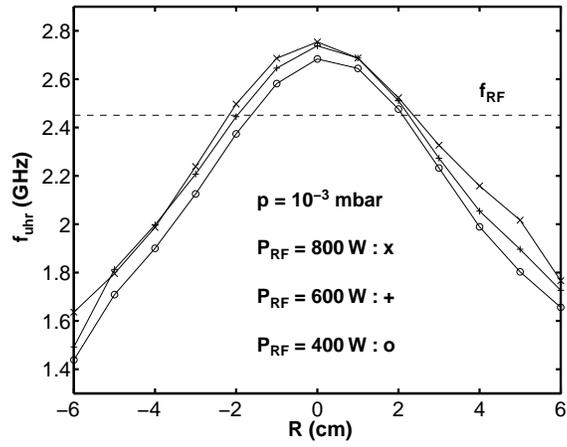,height=2.3in,angle=0}}}
\caption{$f_{uhr}$ with input microwave power variation in hydrogen
plasma.}
\label{hywuhrv}
\end{figure}

\noindent
ECR plasma is produced in the experimental system with hydrogen as the
filling gas at an operating pressure of $1 \times 10^{-3}$ mbar and an
input microwave power of 800 W. The maximum plasma density, $n_e$ in
the system is $3.1 \times 10^{10} ~cm^{-3}$ at the geometrical axis of
the system. Typical plasma temperature, $n_e$ in the system is 12 eV.
Radial profiles of $f_{ce}$, $f_{pe}$ and $f_{uh}$ are calculated and
plotted for the measured density and magnetic field profiles as shown
in figure \ref{wcpuhrhy}. Plasma density during the discharge for
microwave power of 800 W is such that the UHR layer during the
experiment lies at R = 2 cm. The $f_{uh}$ variation with input power is
plotted in figure \ref{hywuhrv} which shows that despite the change in
input power, the UHR layer postion remains unchanged.

\begin{figure}[h]
\centerline{\hbox{\psfig{file=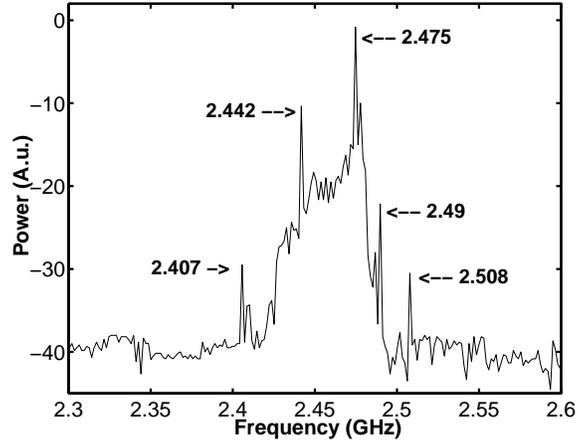,height=2.3in,angle=0}}}
\caption{Parametric decay spectrum in hydrogen plasma.}
\label{hypid}
\end{figure}

\begin{figure}[h]
\centerline{\hbox{\psfig{file=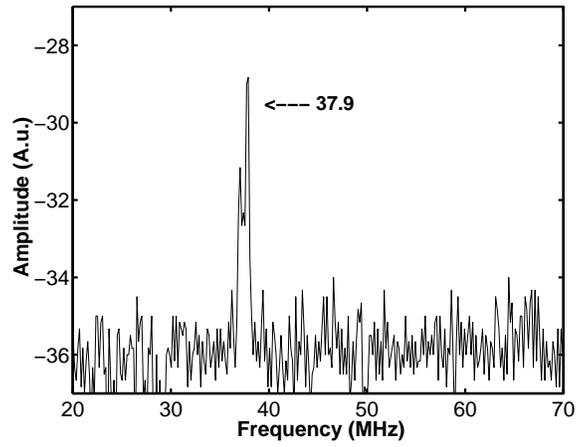,height=2.3in,angle=0}}}
\caption{The residual low frequency.}
\label{lowfp}
\end{figure}

\begin{figure}
\centerline{\hbox{\psfig{file=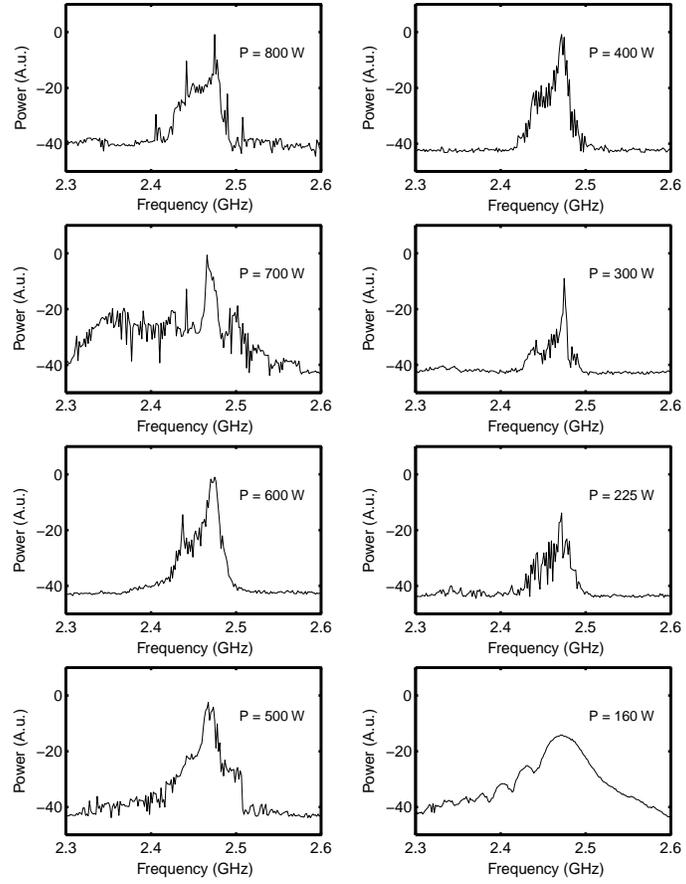,height=4.6in,angle=0}}}
\caption{Parametric decay spectrum with input microwave power
variation.}
\label{pwpids}
\end{figure}

\noindent
The parametric decay spectrum at UHR layer is shown in figure
\ref{hypid}. The residual LH frequency is observed separately as shown
in figure \ref{lowfp} which is calculated as 23.4 MHz hence, it matches
well with the measured peaks. The capacitive probe is moved in hydrogen
plasma from R = 0 cm towards the UHR layer at R = 2 cm. Figure
\ref{pwpids} shows the capacitive probe output at the UHR layer as a
function of microwave power that is increased gradually from 160 W to
the maximum possible value of 800 W. As the power is increased, the
spectrum is broadened on the lower frequency edge and at 300 W a peak
starts appearing at 2.442 GHz which develops as the power is increased,
Power threshold for excitation of the sideband is measured and the peak
at 2.442 GHz appears only beyond 500 W of input power \cite{vipin2}.

\begin{figure}
\centerline{\hbox{\psfig{file=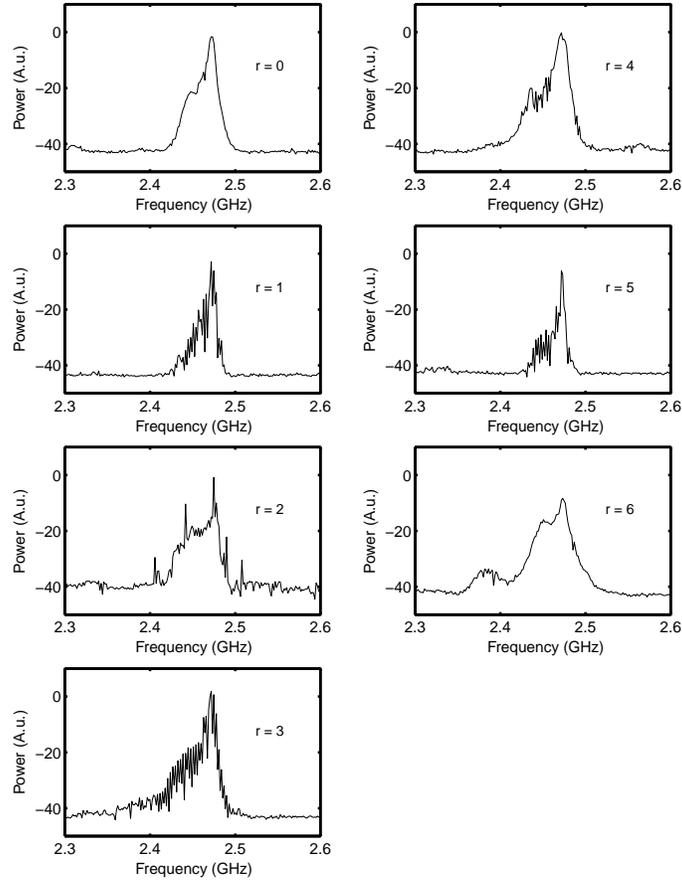,height=4.6in,angle=0}}}
\caption{Radial variation of parametric decay spectrum at 800 W input
microwave power.}
\label{pospow}
\end{figure}

\noindent
At the peak power, the signal is measured along the radius. As the
capacitive probe is placed close to thr UHR layer, within R = 1 cm to R
= 2 cm, the original spectrum splits to smaller peaks as shown in figure
\ref{pospow}. At UHR, the peak at f = 2.442 GHz is distinct. The figure
indicates the two ECR surfaces at R = 0 and R = 6 where the frequency
spectrum is clean without fluctuations. The spectrum at R =2 (UHR
surface) shows the side-peaks with the launched frequency indicating
the parametric decay there. The frequency spectrum at the radial
positions near UHR surface, R = 1 and R = 3 contains fluctuations
suggesting the splitting of power at UHR and propagation in both
forward and reverse directions \cite{vipin2}.

\noindent
The EBW is detected and characterized with two identical capacitive
probes \cite{vipin3}. The axial capacitive probe is fixed at R = 0 cm
and the radial capacitive probe is moved upto UHR layer. This small gap
of 2 cm between the UHR and ECR layers limits the spatial resolution.
The EBW parameters are given in the table below.

\begin{center}
\begin{tabular}{|c|c|c|}
\hline
S.No. & Parameter & Value \\\hline
1. & Gas & Hydrogen \\\hline
2. & $f_0$ GHz & 2.45 $\pm$ 0.02 \\\hline
3. & $f_{EBW}$ GHz & 2.438 \\\hline
4. & $\lambda_B$ mm & 6.56 \\\hline
5. & $v_{\phi}$ $ms^{-1}$ & $1.6 \times 10^7$ \\\hline
6. & $v_g$ $ms^{-1}$ & $3.88 \times 10^4$ \\\hline
7. & $n_{\perp}$ & 20 \\ \hline
\end{tabular}
\end{center}

\begin{figure}
\centerline{\hbox{\psfig{file=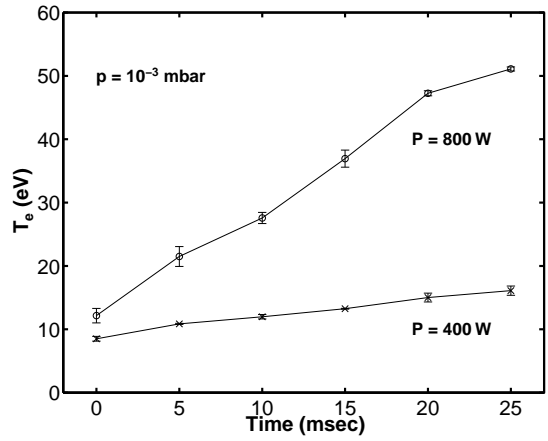,height=2.3in,angle=0}}}
\caption{Time evolution of plasma temperature in hydrogen plasma.}
\label{tetpow}
\end{figure}

\noindent
The mode converted EBW propagate towards the centre of the experimental
system near which the first ECR layer is residing. The EBW is absorbed
at the ECR surface and gives rise to the localized electron heating.
This is evident from the time evolution of plasma temperature as shown
in figure \ref{tetpow} where the temperature increases with time and
then attains a saturated value $\approx$ 50 eV afterwards for 800 W of
input power. This indicates the absorption of the launched EM wave in
the system. The localized heating of plasma with mode converted EBW in
this experimental plasma system is further verified by the observation
of the emission of soft x-rays \cite{vipin4}.

\begin{figure}[h]
\centerline{\hbox{\psfig{file=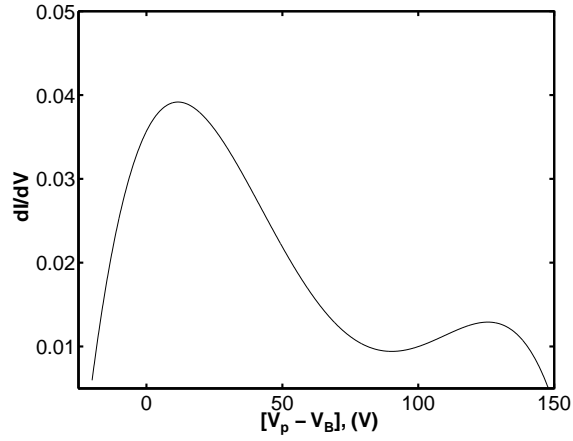,height=2.3in,angle=0}}}
\caption{The electron energy distribution function (EEDF).}
\label{eedf}
\end{figure}

\noindent
The EEDF obtained from the Langmuir probe characteristics at the centre
of the plasma chamber is shown in figure \ref{eedf}. Here, ($V_p$ -
$V_B$) is the accelerating voltage for the electrons in the plasma and
dI/dV signifies the number of such electrons. In the figure, the hump
in the tail of the distribution indicates the presence of high energy
electrons with energies more than 100 eV. These energetic electrons hit
metal targets of tungsten, molybdenum, etc. in the plasma and leads to
emission of soft x-ray. X-ray radiation from plasma can be energy
resolved by using filters in the form of thin Aluminium foils.

\begin{figure}[h]
\centerline{\hbox{\psfig{file=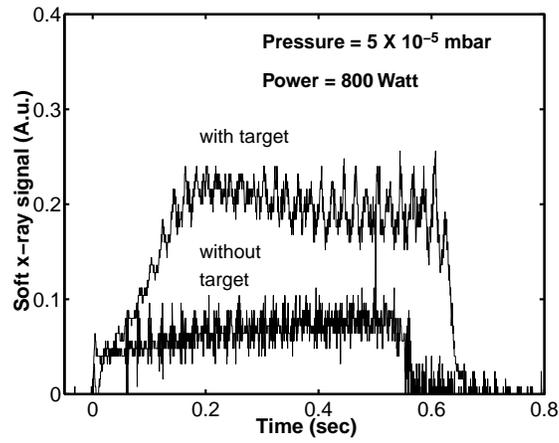,height=2.3in,angle=0}}}
\caption{Soft x-ray signal with and without metal target in plasma.}
\label{sigwpwop}
\end{figure}

\begin{figure}[h]
\centerline{\hbox{\psfig{file=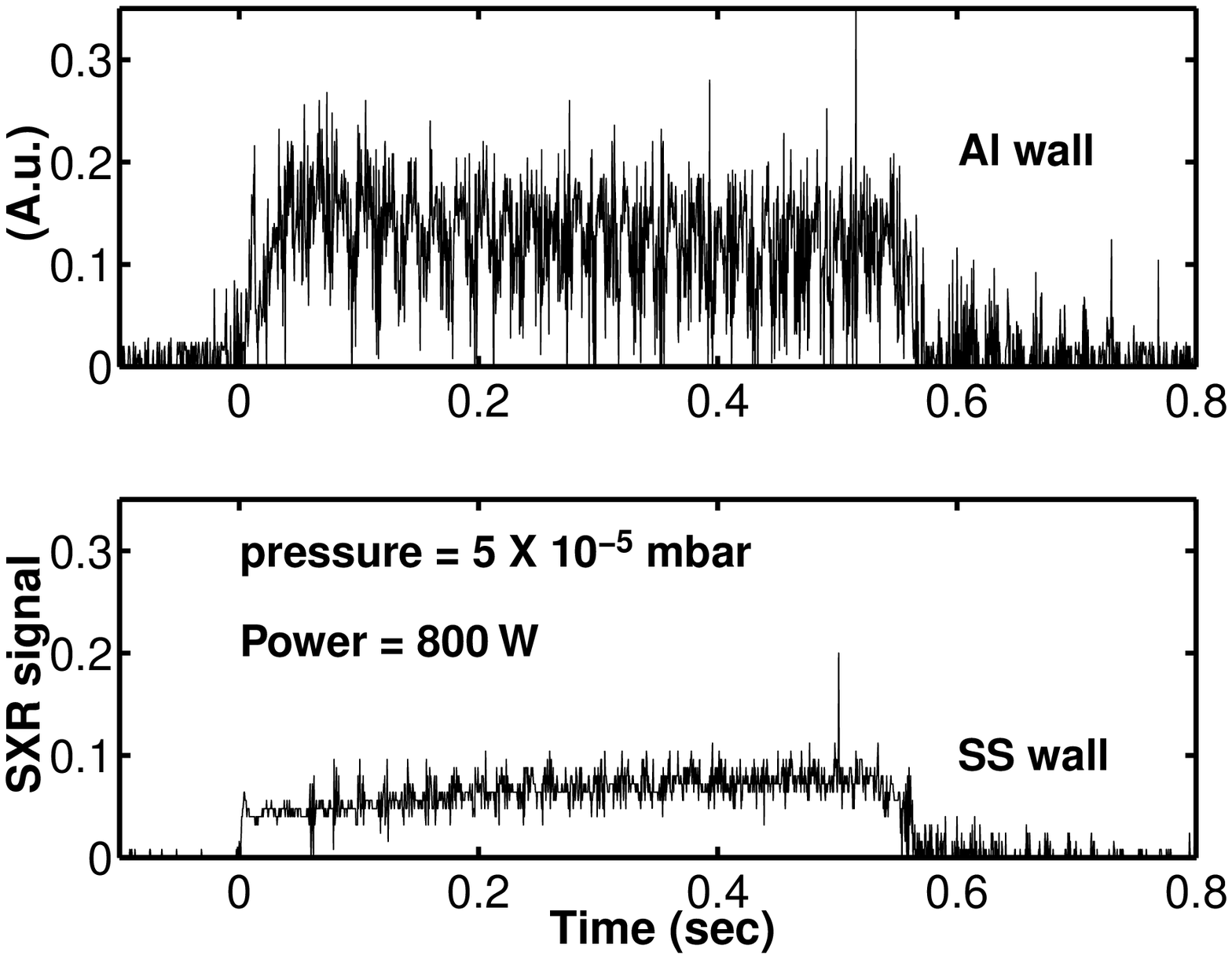,height=2.3in,angle=0}}}
\caption{Soft x-ray signal variation with Al and SS wall.}
\label{alss}
\end{figure}

\noindent
In this experiment also the soft x-ray emission due to the generation
of energetic electrons by the absorption of mode converted EBW is
observed \cite{vipin4}. The soft x-ray emission signal with and without
any target inside the plasma is shown in figure \ref{sigwpwop}. This
signal is very feeble compared to the one with Langmuir probe tip made
of thoriated tungsten placed in the plasma. This indicates that the
presence of a metal target increases the soft x-ray emission. The soft
x-ray emission also depends on the nature of the metal target as shown
in figure \ref{alss} where the signals for both the targets differ.

\section{Conclusion}

\noindent
Plasma heating due to X-B mode conversion is observed in a linear ECR
plasma system. The X-B mode conversion is nonlinear in nature associated
with parametric decay of the launched wave frequency. Parametric decay
spectrum of the launched X-mode is observed at the UHR layer in system
with capacitive probe. The decay spectrum contains $\omega_0 \pm
\omega_{LH}$ about the launched frequency, $\omega_0$ (2.45 + 0.02 GHz)
indicating the three-wave nonlinear interaction. The experimentally
observed value of $\omega_{LH}$ in the spectrum matches quite well with
the estimated values. The parametric decay spectrum has an input power
threshold ($\approx$ 600 W) to occur. This is in accordance with the
estimated input power values. The parametric decay peaks are observed
only near the calculated UHR layer in the system. The shift in probe
position from UHR vanishes the peaks. The EBW generated near the UHR
surface as a result of the mode conversion is experimentally observed
and characterized. The EBW then propagates towards the first harmonic
cyclotron resonance layer residing near the centre of the geometrical
axis of the system and is absorbed. This absorption leads to the
localized electron heating at that position. Experimental results
indicate the presence of high energy electrons. This is further verified
by EEDF obtained from the I-V characteristic at the centre of the
system. The soft x-ray emission is observed from the ECR plasma using a
VPD. The soft x-ray signal is feeble in the absence of a target and gets
enhanced in the presence of a tungsten target. The signal is also
observed to be a dependent function of the material of the target. 

\newpage

\newpage

\begin{figure}[h]
\centerline{\hbox{\psfig{file=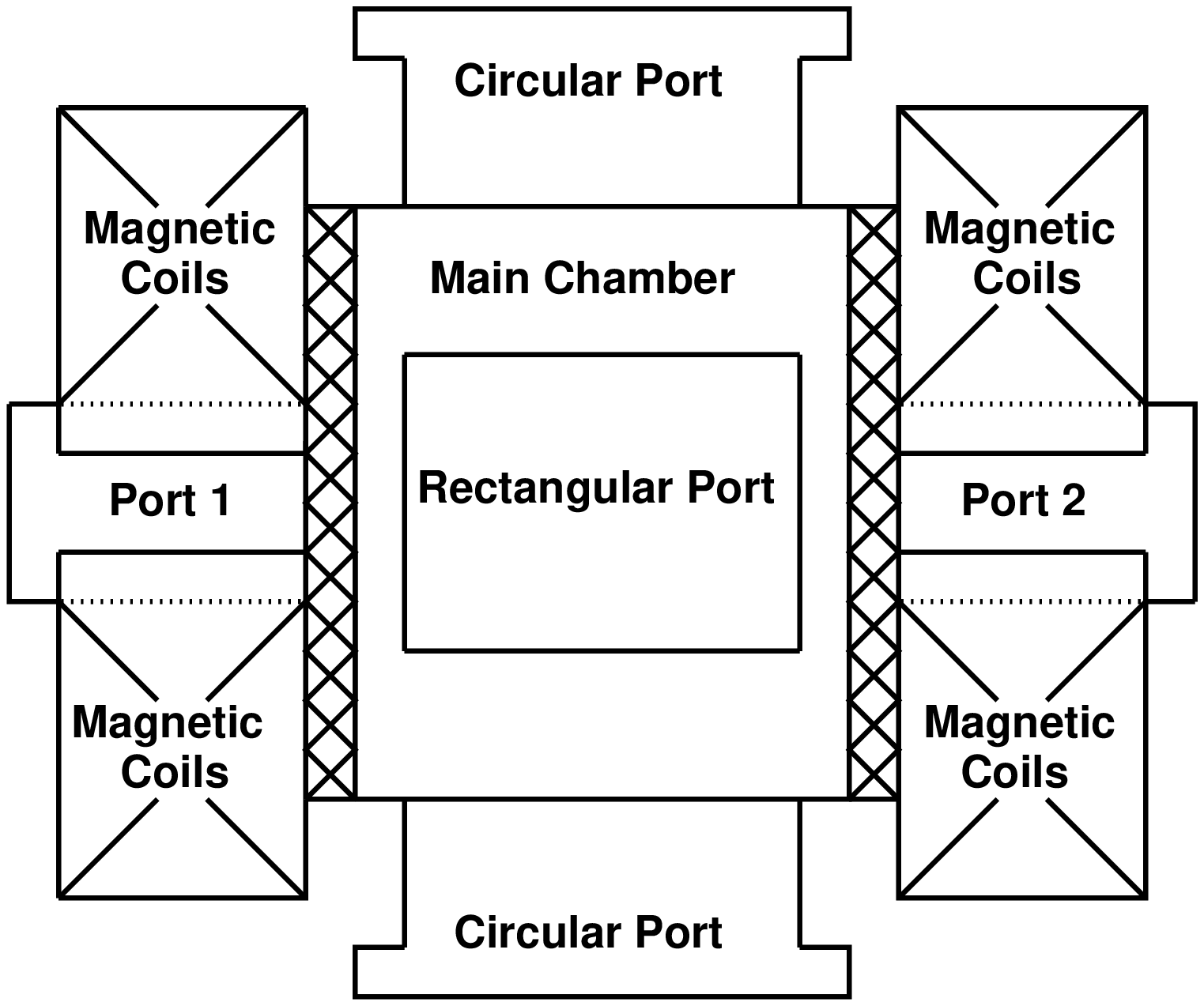,height=2.3in,angle=0}}}
\caption{Main chamber.}
\label{syssol}
\end{figure}

\newpage

\begin{figure}[h]
\centerline{\hbox{\psfig{file=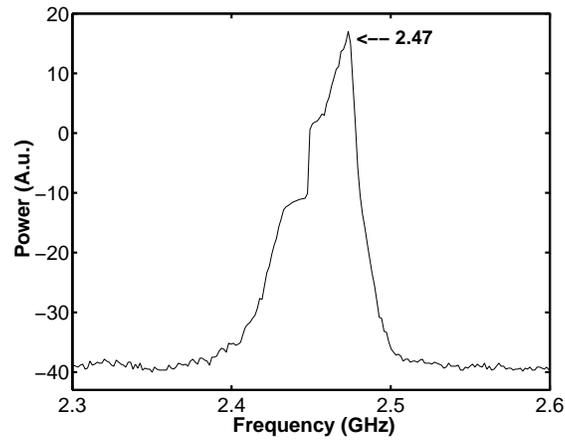,height=2.3in,angle=0}}}
\caption{Power output from magnetron at forward port of directional
coupler.}
\label{magfs}
\end{figure}

\newpage

\newpage

\newpage

\newpage

\begin{thebibliography}{99}
\bibitem[\dagger]{mail1}E-mail address: vipin@ipr.res.in
\bibitem[\sharp]{mail2}E-mail address: dbora@ipr.res.in

\bibitem{lloyd} Brian Lloyd, Plasma Phys. Cont. Fusion, {\bf 40}, A119
(1998)

\bibitem{manheimer} Wallace M. Manheimer, Infrared and Millimeter Waves
Vol. 2 : Instrumentation, (Academic Press, New York, 1979), chapter 5,
299

\bibitem{england} A.C. England, IEEE Trans. Plasma Sci., {\bf PS-12},
124 (1984)

\bibitem{eldridge} O.C. Eldridge and A.C. England, Nucl. Fusion,
{\bf 29}, 1583 (1989)

\bibitem{erckmann} V. Erckmann and U. Gasparino, Plasma Phys. Cont.
Fusion, {\bf 36}, 1869 (1994)

\bibitem{abrakov} V.V. Abrakov, D.K. Akulina, Eh. D. Andryukhina, {\it
et. al.}, Nucl. Fusion, {\bf 37}, 233 (1997)

\bibitem{porkolab} M. Porkolab, L. Friedland and I.B. Bernstein, Nucl.
Fusion, {\bf 21}, 1463 (1981)

\bibitem{bernstein} Ira B. Bernstein, Phys. Rev., {\bf 109}, 10 (1958)

\bibitem{piliya} A.D. Piliya, A. Yu Popov and E.N. Tregubova, Plasma
Phys. Control. Fusion, {\bf 45}, 1309 (2003)

\bibitem{stix} Thomas H. Stix, Phys. Rev. Lett., {\bf 15}, 878 (1965)

\bibitem{istomin} Ya. N. Istomin and T.B. Leyser, Phys. Plasmas,
{\bf 2}, 2084 (1995)

\bibitem{sugai} H. Sugai, Phys. Rev. Lett., {\bf 47}, 1899 (1981)

\bibitem{nakajima} S. Nakajima and H. Abe, Phys. Rev. A, {\bf 38}, 4373
(1988)

\bibitem{ram} A.K. Ram and S.D. Schultz, Phys. Plasmas, {\bf 7}, 4084
(2000)

\bibitem{mcdermott} F.S. McDermott, G. Bekeffi, K.E. Hackett, J.S.
Levine and M. Porkolab, Phys. Fluids, {\bf 25}, 1488 (1992)

\bibitem{laqua1} H.P. Laqua, H. Maassberg, N.B. Marushchenko, F. Volpe,
A. Weller, W7-AS Team, W. Kasparek and ECRH-Group, Phys. Rev. Lett.,
{\bf 90}, 075003-1 (2003)

\bibitem{laqua2} H.P. Laqua, V. Erckmann, H.J. Hartfu$\beta$, H. Laqua,
W7-AS Team and ECRH group, Phys. Rev. Lett., {\bf 78}, 3467 (1997)

\bibitem{gekelman} W. Gekelman and R.L. Stenzel, Rev. Sci. Instrum.,
{\bf 46}, 1386 (1975)

\bibitem{vipin1} Vipin K. Yadav and D. Bora, Pramana, {\bf 63}, 563,
(2004)

\bibitem{langmuir} I. Langmuir, Phys. Rev., {\bf 36}, 954 (1929)

\bibitem{vipin2} Vipin K. Yadav and D. Bora, Phys. Plasmas, {\bf 11},
3409 (2004)

\bibitem{vipin3} Vipin K. Yadav and D. Bora, Phys. Plasmas, {\bf 11},
4582 (2004)

\bibitem{rao} C.V.S. Rao, Y. Shankara Joisa and C.J. Hansalia, A.K.
Hui, Ratan Paul and Prabhat Ranjan, Rev. Sci. Instrum., {\bf 68}, 1142
(1997)

\bibitem{vipin4} Vipin K. Yadav and D. Bora, Plasma Sources Sci.
Technol., {\bf 13}, 231, (2004)

\end{thebibliography}
\end{document}